\documentclass[]{spie}

\usepackage{amsmath,amsfonts,amssymb}
\usepackage{jour}
\usepackage{graphicx}
\usepackage[colorlinks=true, allcolors=blue]{hyperref}
\usepackage{bm}

\title{Computation of the Fresnel diffraction of starshades based on a polygonal approximation}

\author{Simon Prunet}
\author{Claude Aime}
\author{André Ferrari}
\author{Céline Theys}

\affil{Universit\'e C\^ote d'Azur, Observatoire de la C\^ote d'Azur, CNRS, Nice, France}

\authorinfo{Further author information: Send correspondence to Simon Prunet  (simon.prunet@oca.eu)}

\begin{document}
\maketitle
\begin{abstract}
    The design of starshades, i.e. external occulters for stellar coronography, relies on the fast and precise computation of their associated diffraction patterns of incoming plane waves in the telescope aperture plane. We present here a method based on a polygonal approximation of the occulter's shape, that allows fast computation of their diffraction patterns in the Fresnel approximation, without aliasing artefacts. It is competitive with respect to methods based on direct 2D Fourier transforms, or Boundary Diffraction Wave algorithms.
\end{abstract}

\section{Introduction}
\label{sec: Introduction}

The goal of starshade screens is to obtain, with the help of an external screen placed in the line of sight of a star, a zone of shadow of a size large enough to cover the entrance pupil of a large telescope, while at the same time sustaining a small enough angle on the sky in order to leave extrasolar planets orbiting that star visible. The precise shape of the occulter is optimized in order to obtain high contrast shadows on the entrance pupil, typically on the order of $10^{-10}$ in intensity, with respect to the unobscured region. 

To reach such high contrasts, computations of the occulter's diffraction of light waves from the star need to be achieved with great accuracy, and with sufficient speed to be able to efficiently explore the optical system's parameters (e.g. distance to the telescope, inner and outer radii of the occulter, number and shape of the petals, robustness to phase errors, etc.), as well as other sources of parasitic light (e.g. solar and zodiacal glare). State of the art diffraction computations, used e.g. in the SISTER\cite{hildebrandt2021} simulation suite, rely on Boundary Diffraction Wave algorithms\cite{Cady2012,miyamoto1962}. The major advantage of the latter methods, as compared to classical Fresnel convolution integrals based on 2D Fourier transforms, is that they rely on boundary, 1D integrals to compute the diffraction patterns, achieving very substantial gains in terms of computation speed. However, computations need to be redone for each wavelength, with no possibility to separate computations related to the occulter shape and to light diffraction per se.

We describe here a method that is following the traditional approach of 2D Fresnel convolution integrals, but with the advantages of Boundary Diffraction methods, in the sense that at least part of the computations are done using 1D boundary integrals. This is achieved by assuming that the occulter can be described by a polygon, which can be done with arbitrary precision by increasing the number of vertices. In \autoref{sec: Formalism}, we describe the formalism of the method. In \autoref{sec: Vertices}, we discuss the choice of vertices on the occulter's perimeter, and in \autoref{sec: Results}, we show how this method converges in terms of diffracted profiles for the NW2\cite{hildebrandt2021} optimized starshade, and illustrate how it compares to traditional Fresnel convolutions. In \autoref{sec: timings}, we discuss the scaling of the computation time with respect to both the number of vertices, and the number of output spatial frequencies of the Fourier transform of the occulter's mask, and finally conclude in \autoref{sec: Conclusions}. More details can be found in Ref.~\citenum{aime2024}.

\section{Formalism}
\label{sec: Formalism}
For the NW2 occulter of typical radius $R=36\, \mathrm{m}$, at a distance $Z=119770\, \mathrm{km}$ from the aperture plane, the Fresnel number $\Phi = R^2/(\lambda Z) \simeq 21.6 $
at $\lambda = 500\,\mathrm{nm}$, which places the diffraction well within the Fresnel limit.
Let us consider an occulting mask $f(\bm{r}),\,\bm{r}=(x,y)$, we have $f(\bm{r})=1 -t(\bm{r})$, where $t(\bm{r})$ is a transparent mask that is spatially bounded at radius $R$, i.e. $t(||\bm{r}||)=0$ for $||\bm{r}||>R$. Following Ref.~\citenum{aime2024}, the diffraction amplitude by an occulter at distance $Z$ from the telescope aperture located at $z=0$, from a normal incident wave (incidence angle $\bm{\xi}={\bm 0}$), is:
\begin{align} \label{eq:FresnelConvolution}
\Psi_{\lambda }(\bm{0},\bm{r},Z) &=f(\bm{r})   * \frac{1}{i \lambda Z}\exp(i \pi \frac{||\bm{r}||^2}{\lambda Z}) \nonumber \\
&=1-t(\bm{r})   * \frac{1}{i \lambda Z}\exp(i \pi \frac{ ||\bm{r}||^2}{\lambda Z}).
\end{align}
For incoming waves with an angle $\bm{\xi} \neq \bm{0}$, we can deduce the diffraction pattern in the following way:
\begin{equation}
  \Psi_{\lambda }(\bm{\xi},\bm{r},Z) =\Psi_{\lambda}(0,\bm{r}+\bm{\xi} Z,Z)
   \exp(- 2 i \pi \frac{\bm{r}.\bm{\xi}}{\lambda} ).
\end{equation}
The 2D Fresnel convolution in \autoref{eq:FresnelConvolution} can be computed using Fourier transforms:
\begin{equation}\label{eq:FourierConvolution}
\Psi_{\lambda }(\bm{0},\bm{r},Z)  =1- \mathcal{F}^{-1}[\hat{t}(\bm{\rho})\times \mathcal{F}[\frac{1}{i \lambda Z}\exp(i \pi \dfrac{ ||\bm{r}||^2}{\lambda Z})]],
\end{equation}
where $\hat{t}(\bm{\rho})=\mathcal{F}[t(\bm{r})]$ is the Fourier transform of the transmission profile, and $\bm{\rho}=(u,v)$ is the vector of spatial frequencies. We see that the wavelength only enters in the quadratic phase term, and the Fourier transform of the transparent mask $\hat{t}(\bm{\rho})$ can be computed just once for all wavelengths.
It is interesting to note that Fourier transforms present in the right hand side of \autoref{eq:FourierConvolution} have very different properties: the first one is the Fourier transform of a binary transparency mask, with sharp boundaries, and therefore has slowly decreasing power at large $||\bm{\rho}||$, while the second, because of the finite extent of the field, has power localized at low spatial frequencies\cite{aime2024}. 

While the latter means that the Fourier product in \autoref{eq:FourierConvolution} only needs to be done for a relatively small number of discrete values of $\bm{\rho}$ before being transformed back to image space via an inverse FFT, the former means that great care must be taken to avoid aliasing while computing the Fourier transform $\hat{t}(\bm{\rho})$. 

To address this last problem, direct approaches using 2D pixelized versions of the occulter necessitate very high resolutions: typically, in the case of the NW2 setup, up to $2^{23}$ pixels per dimension\cite{aime2024}, to achieve the contrast at the center of the field predicted by analytical means\cite{Vanderbei2003}. Therefore, we would like to use a method that retains the computational advantages of Boundary Wave Diffraction techniques, while retaining the separation between the occulter dependent term and the wavelength dependent Fresnel kernel. 

First introduced in the context of radio receivers\cite{Lee1983} and later revisited and further developed in the field of neutron and X-ray scattering\cite{Wuttke2021}, close forms of continuous 2D Fourier transforms of indicator functions of polygonal shapes are expressed as discrete sums over polygon vertices with geometric and phase weights. These transforms, being continuous, are by construction free of aliasing artefacts, and their precision is only limited by the quality of the polygonal approximation of the occulter, that can be controlled by the number and position of the polygon vertices.

Let us call $\bm{v}_j,\, j=1...N$ the coordinate vectors of the polygon vertices, $\bm{r}_j = (\bm{v}_{j+1}+\bm{v}_j)/2$ the middle of edge $j$, and $\bm{e}_j = (\bm{v}_{j+1}-\bm{v}_{j})/2$ the vector joining $\bm{v}_j$ and $\bm{r}_j$. The continuous 2D Fourier transform of the indicator function of the polygon can, due to Stokes' theorem, be expressed as a contour integral on the polygon perimeter. Following Wuttke\cite{Wuttke2021}, let us define the function $\mathbf{g}(\bm{r}) = 2\pi \hat{\mathbf{n}}\times\bm{\rho}e^{i2\pi\bm{\rho.r}}$, where $\bm{a}\times\bm{c}$ is the cross-product of vectors $\bm{a}$ and $\bm{b}$. Stokes' theorem reads:
\begin{equation}
\iint_\Gamma \mathrm{d}^2r \,\hat{\mathbf{n}}.(\nabla\times \mathbf{g}) = \oint_{\partial\Gamma}\mathrm{d}\mathbf{r} .\mathbf{g},
\label{eq:stokes}
\end{equation}
with $\bm{a.b}$ the scalar product of vectors $\bm{a}$ and $\bm{b}$, $\Gamma$ the polygonal shape and $\partial\Gamma$ its oriented perimeter.

In our case, the left hand side of \autoref{eq:stokes} is equal to:
\begin{align}
    \iint_\Gamma \mathrm{d}^2r \,\hat{\mathbf{n}}.(\nabla\times \mathbf{g}) 
    &= 4\pi^2\hat{\mathbf{n}}.[i\bm{\rho}\times(\hat{\mathbf{n}}\times\bm{\rho}]\iint_\Gamma \mathrm{d}^2r\, e^{i2\pi\bm{\rho.r}} \nonumber\\
    &= 4\pi^2 i|\hat{\mathbf{n}}\times\bm{\rho}|^2 \hat{t}(\bm{\rho}) \nonumber\\
    &= 4\pi^2 i ||\bm{\rho}||^2 \hat{t}(\bm{\rho}),
    \label{eq:stokes-double}
\end{align}
while the right hand side can be expressed as a sum of integrals over the polygon edges. Parametrizing the points of a polygon edge by $\bm{r}_j(\lambda) = \bm{r}_j + \lambda\bm{e}_j$, the right hand side reads:

\begin{align}
    \oint_{\partial\Gamma}\mathrm{d}\mathbf{r.g} &= \sum_{j=1}^N\mathbf{e}_j.\int_{-1}^{1}\mathrm{d}\lambda\mathbf{g}(\mathbf{r}_j(\lambda)) \nonumber\\
    &= 2\pi\sum_{j=1}^N \left[\mathbf{e}_j,\hat{\mathbf{n}},\bm{\rho}\right] \int_{-1}^{1}\mathrm{d}\lambda e^{i2\pi\bm{\rho.r}_j(\lambda)} \nonumber\\
    &= 4\pi\sum_{j=1}^N \left[\hat{\mathbf{n}},\bm{\rho},\mathbf{e}_j\right]\mathrm{sinc}(2\pi\bm{\rho.e}_j)e^{i2\pi\bm{\rho.r}_j}.
    \label{eq:stokes-single}
\end{align}
Using \autoref{eq:stokes-double} and \autoref{eq:stokes-single}, we finally get:
\begin{equation}\label{eq:tf-poly}
\hat{t}(\bm{\rho}) = \frac{1}{i\pi ||\bm{\rho}||^2} \sum_{j=1}^{N} \left[ \hat{\mathbf{n}},\bm{\rho},\mathbf{e}_j \right] 
\mathrm{sinc}(2\pi\bm{\rho.e}_j)e^{i2\pi\bm{\rho.r}_j}.
\end{equation}
This equation expresses the continuous Fourier transform of the occulting mask, as a sum over the polygon vertices with simple geometric and phase weights. We emphasize the fact that no discretization has been necessary up to now, so that this formula is free of any aliasing artefacts. We note that the sum runs over the polygon vertices, a discrete set of points on the mask boundary, which number therefore scales as the square root of the number of pixels in a 2D sampled version of the mask, as in Boundary Diffraction methods.

\section{Choice of polygon vertices}
\label{sec: Vertices}

Some care needs to be used when choosing the polygon vertices to obtain a good approximation of the petal occulter's shape. Let us recall how the mask is designed in the first place; it is itself a binary approximation of a circularly symmetric intensity filter whose (variable) transmission profile is optimized to obtain a required intensity contrast level on the telescope aperture\cite{Vanderbei2007}. The Fresnel convolution kernel itself being circularly symmetric (see \autoref{eq:FresnelConvolution}), the diffraction pattern at the center of the field will be the same as that of the symmetric, variable intensity filter, provided that the azimuthally averaged radial profile of the binary mask is identical to that of the variable filter. 

\begin{figure}[!ht]
\includegraphics[width=\textwidth]{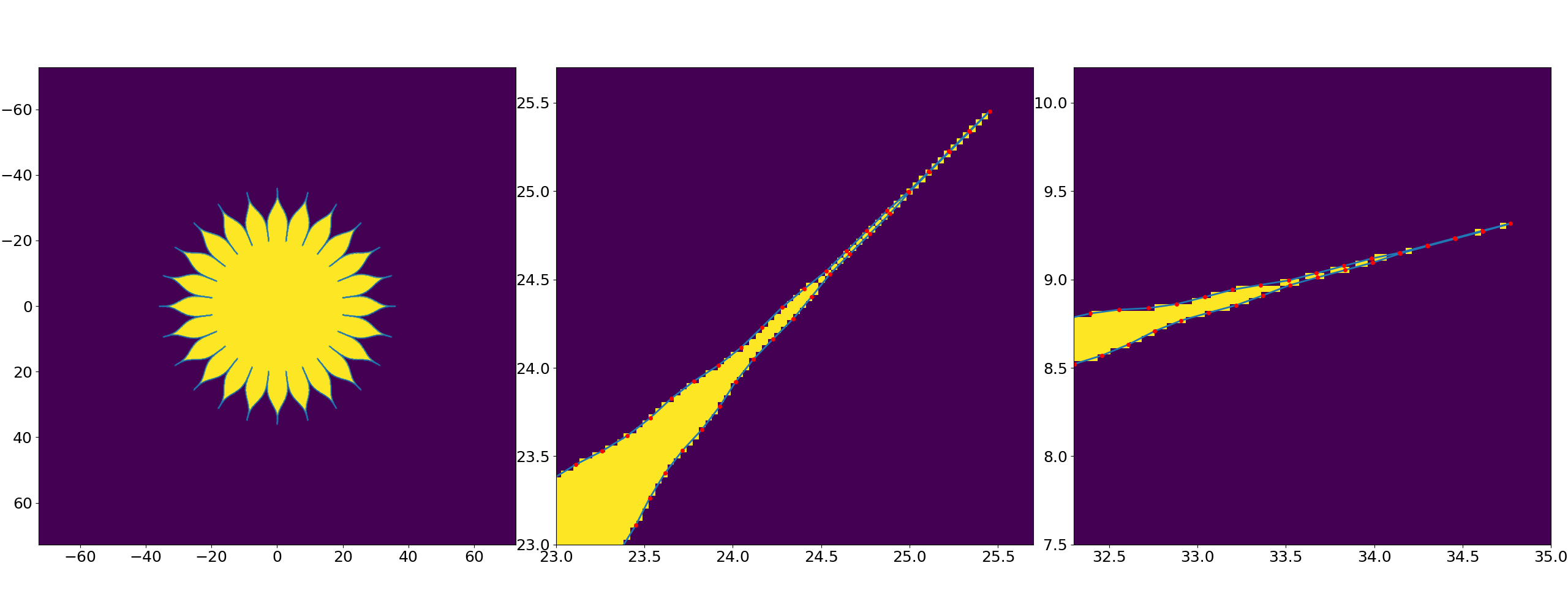}
\caption{Pixelized petal mask of the NW2 setup, with $24$ petals (yellow), together with a polygonal approximation of the continuous mask (blue line) and the associated vertices (red points). There are $100$ vertices per half petal, for a total of $4752$ vertices in total, comparable to the square root of the number of pixels. \emph{Left}, \emph{middle} and \emph{right} panels respectively show the entire mask with it padding region, and zooms on the petal centered around $45^\circ$ and $15^\circ$ degree of azimuth. These numbers are chosen for display purposes only, as the actual numbers chosen in the diffraction calculations will be much larger to achieve the required contrast levels. Dimensions are in meters.\label{fig: multi-panels-mask}}
\end{figure}

At radii $r \in[r_{\mathrm{in}},r_{\mathrm{out}}]$ for which the variable filter profile transmission $f(r)$ is between $0$ and $1$, this can be achieved by introducing a number $P$ of petal shapes, that are deduced from a single one by rotations of angle $2\pi/P$.
The shape of the reference petal is then chosen such\cite{Vanderbei2003} that 
\begin{equation}\label{eq:petal-sampling}
\int_{-\theta(r)}^{\theta(r)}\mathrm{d}\theta = \frac{2\pi}{P} f(r), 
\end{equation}
with $r,\theta(r)$ defining the polar coordinates of the boundary of the petal upper half ($0\leq\theta(r)\leq \pi/P)$, the lower half being deduced from the latter by symmetry with respect to the horizontal axis.

For the polygon vertices, we made the simple choice of doing a regular sampling of the upper half reference petal that is linear in radius, with $n$ points on that interval. We will explore later the effect of varying $n$ on the intensity contrast achieved by the polygonal mask thus defined. In \autoref{fig: multi-panels-mask}, we show a pixelized mask and its border as defined by the polygonal approximation. In this example, the linear pixel size is $4096$ (including the padding region needed for the FFT), while the total number of vertices of the polygon is $4752$, corresponding to $n=100$ vertices per half petal. We see that, with a number of polygon vertices roughly equal to the square root of the number of pixels, we obtain a visually smoother approximation of the continuous occulter shape.
In the next section, we will show the results obtained for the radial intensity contrasts with the polygonal approximation, as a function of the number of vertices.

\section{Results}
\label{sec: Results}

\begin{figure}[!ht]
\begin{center}
\includegraphics[width=\textwidth]{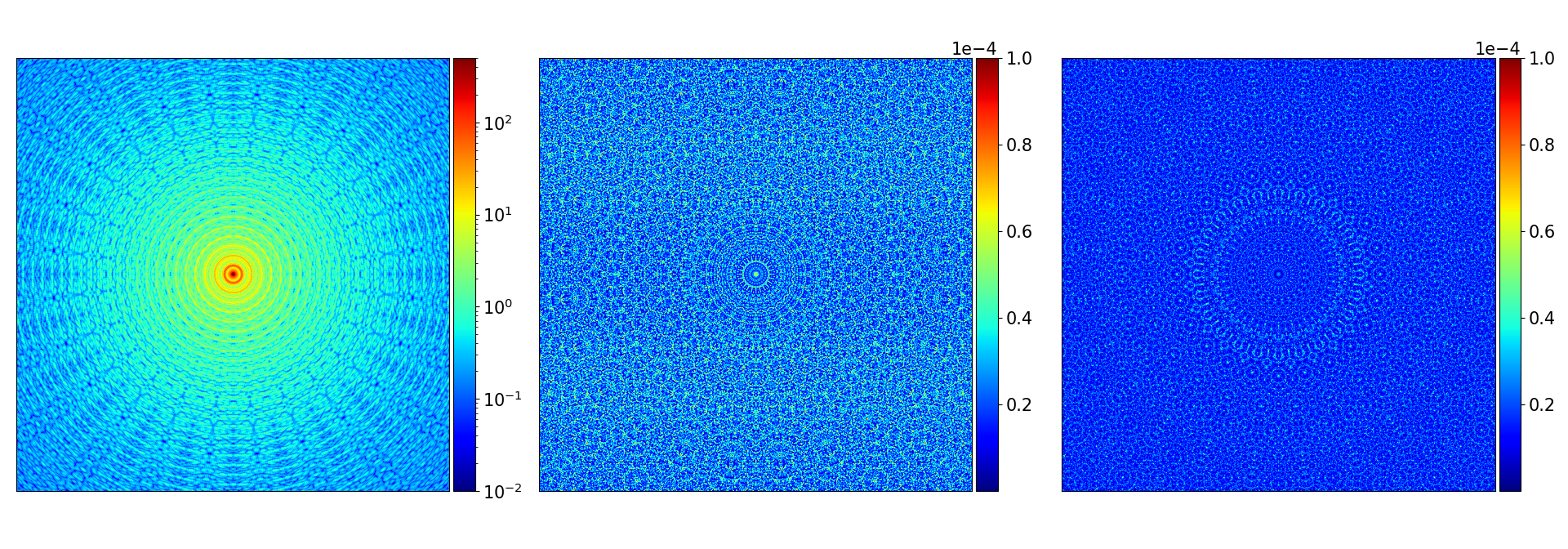}
\end{center}
\caption{\label{fig: TF masks} 
\emph{Left panel}: Central $2000\times2000$ pixel regions of the petal occulter 2D Fourier transform, computed at ($8192\times8192$) discrete locations using the continuous Fourier transform of the polygonal approximation, with $8000$ vertices per half petal, for a total of $383952$ vertices. \emph{Middle panel}: Difference between the Fourier transform computed with the polygonal approximation and the 2D FFT of a pixelized mask of very high resolution ($2^{23}\times2^{23}$, which includes a padding zone three times larger than the occulter itself to reduce aliasing). \emph{Right panel}: Difference between two transforms using both the polygonal approximation, with $4000$ and $8000$ vertices per half petal, respectively.
}
\end{figure}

Still using the NW2 setup, \autoref{fig: TF masks} compares the computation of the 2D Fourier transform of the masks, in their respective approximations as pixelized image and polygonal shape. In this comparison, the number of pixels  \emph{per axis} of the discrete mask is $2^{23}=8388608$, while the occulter support is limited to $2^{21}=2097152$ pixels per axis, the large extra padding zone being used to control aliasing in the FFT. The number of polygonal vertices is set to $n=8000$ per half petal, for a total of $383952$ vertices. Note that only the real part of the Fourier transform is shown, as the imaginary part is null by symmetry. The left panel shows the mask transform obtained with the polygonal approximation, the middle panel shows the difference of the continuous Fourier transform using the polygonal approximation and that obtained from the 2D FFT of the pixelized mask, while the right panel shows the difference of Fourier transforms obtained with the polygonal approximation, with respectively $4000$ and $8000$ vertices per half petal.

Amplitude differences in the middle panel are homogeneous on the order of a few $10^{-5}$, while the maximum of the Fourier transform, corresponding to the area in meters of the occulter, is $\sim 2691$. Apart from differences linked to the peak of the Fourier transform at large scales (central frequencies), the difference pattern in the middle panel appears repetitive, with a mashrabiya-like appearance, and is most probably due to aliasing artifacts of the pixelized mask transform. This is further confirmed by the difference map of the right panel between two transforms using the polygonal approximation with different sampling rates: this difference map is typically of a smaller amplitude already compared to that of the middle panel, showing the level of convergence of the polygonal approximation in Fourier space.

\begin{figure}[!ht]
\begin{center}
\includegraphics[]{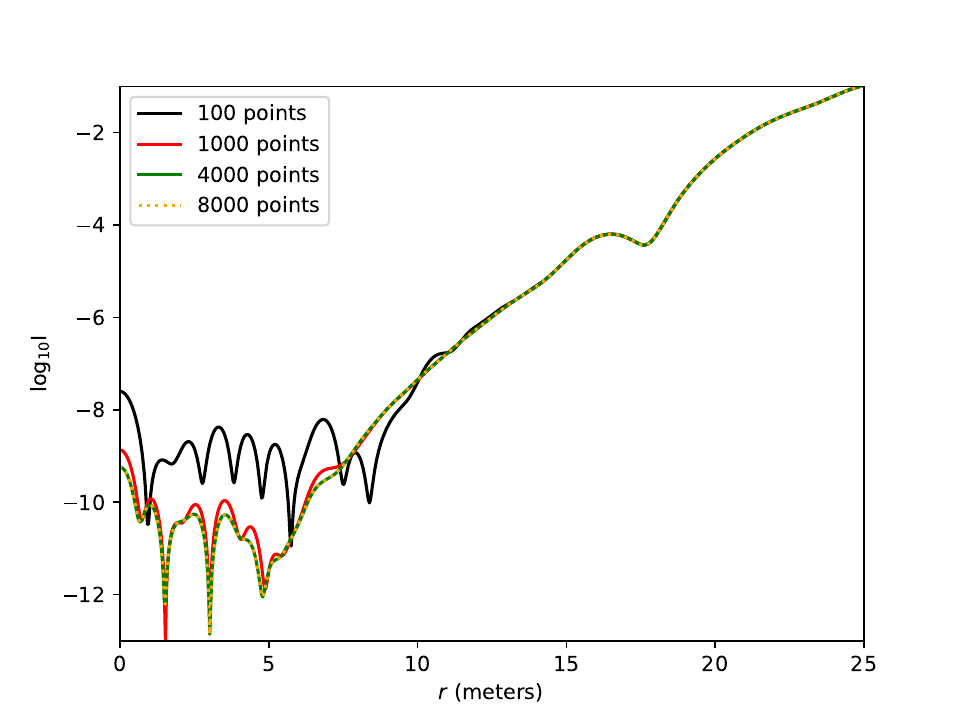}
\end{center}
\caption{\label{fig: number-of-points} Radial profile of intensity contrast of the NW2 occulter diffraction pattern, as a function of the distance to the optical center, in the telescope aperture plane. The number of points per half petal is shown in the legend. We can see that by $n=4000$ points per half petal, the profile computation has converged, as both green ($n=4000$) and dotted orange ($n=8000$) lines are superposed.}
\end{figure}

Another test of the polygonal approximation consists, of course, in computing the numerical properties of the diffraction pattern in the telescope aperture plane, and more specifically the level of intensity contrast achieved using polygonal approximations to the occulter. \autoref{fig: number-of-points} shows, in logarithmic scale, a radial intensity profile for an incoming plane wave of unit amplitude, taken along the horizontal axis. The figure shows the radial profile for different sampling rates of the polygonal approximation, going from $100$ to $8000$ vertices per half petal of the NW2 occulter. We notice that the $10^{-10}$ intensity contrast is achieved within a radius of $5$ meters (except at the very center), and also that the radial profiles for polygonal approximations, respectively based on $4000$ and $8000$ vertices per half petal, are indistinguishable in this figure, suggesting that the polygonal sampling used here is large enough.

Coming back to the sampling strategy discussed in \autoref{sec: Vertices}, we used by simplicity a sampling of polygon vertices that is linear in radius $r$, the corresponding azimuth (for vertices in the first upper half petal) being chosen according to \autoref{eq:petal-sampling}. However, there is no reason \emph{a priori} to think that this sampling choice is optimal, at a fixed sampling rate, in terms of minimizing the errors linked to the polygonal approximation of the continuous occulter. We therefore investigated a different sampling method, based on pruning of a polygon with very high sampling rate (typically around $100k$ vertices per half petal), using the Douglas-Peucker simplification algorithm\cite{douglas-peucker1973}. We stopped the pruning process when we recovered (roughly) the same number of vertices as in the case of linear radius sampling with $8000$ samples per half petal. Surprisingly enough, the results obtained on e.g. the diffraction radial intensity profile were not competitive with the simple, linear sampling in radius with a similar number of vertices. 

However, the Douglas-Peucker algorithm uses a user-specified function for the distance of a point to a given segment, and we haven't explored alternative functions to the default orthogonal projection distance. Another possible direction of improvement, would be to make the sampling policy an integral part of the intensity contrast maximising strategy, by defining a differentiable loss function on the diffraction pattern that could be minimized by allowing sampling points to slide along the continuous petal border, this is however out of the scope of the present work.

\section{Computational aspects}
\label{sec: timings}
We give here a few details about the implementation of the continuous Fourier transform of the polygon indicator function, as expressed by \autoref{eq:tf-poly}. For arbitrary coordinates $\bm{\rho}$ in the frequency plane, \autoref{eq:tf-poly} is implemented in a straightforward manner, either on cpu using \texttt{numpy} operations, or on gpu if available using \texttt{cupy} arrays. It is obvious from this equation that the number of operations scales linearly with respect to both the number of points in frequency space where the transform is needed, as well as the number of vertices in the polygon.

The product of both numbers also defines, up to a constant multiplier, the memory footfprint. Care must thus be taken not to overload the available memory, be it that of the cpu or of the gpu, depending where the computation is done. The calculation is therefore done by slicing the array of required positions in the frequency plane and iterating over that slicing; the user specifying the available memory resources, the number of slices is computed in order to fill at best the available memory, minimizing the number of host to device transfers (in case of a computation made on gpu). 

\begin{figure}
    \centering
    \includegraphics[width=\columnwidth]{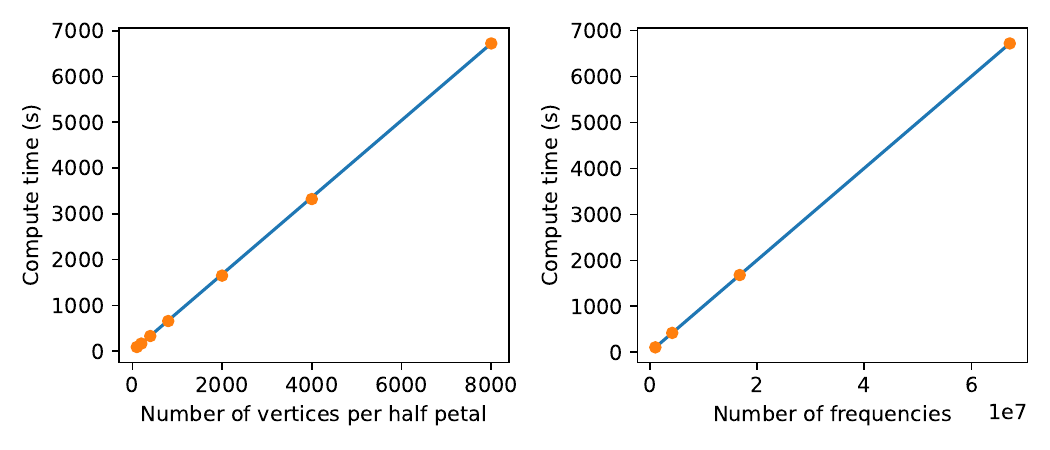}
    \caption{\emph{Left panel}: Compute time of the continuous Fourier transform of the polygonal indicator function, as a function of the number $n$ of vertices per half petal of the NW2 setup. The total number of petals is equal to $N=(n-1)*48$, with $383952$ vertices for $n=8000$. \emph{Right panel}: Same, but as a function of the number of output spatial frequencies. The last point corresponds to a $8192\times8192$ frequency grid. Both timings show the expected linear scaling from \autoref{eq:tf-poly}. }
    \label{fig:timings}
\end{figure}

\autoref{fig:timings} shows the measured scaling (orange points) of the computation time for the NW2 occulter with its $24$ petals, when varying either the sampling of the petals (left panel) or the number of output frequencies (right panel), obtained here when using a Tesla V100 GPU. Both follow the expected linear scaling.

\section{Conclusions} \label{sec: Conclusions}
We presented a new way of computing the Fresnel diffraction pattern in the aperture plane of a telescope, caused by a binary occulter with $24$ petals in the NW2 setup\cite{hildebrandt2021}. It is based on a traditional 2D convolution with Fresnel kernels, however, it presents important differences with the method based on FFTs of the discretized problem. Indeed, this method is based on a polygonal approximation to the occulter's perimeter, allowing the computation of the continuous 2D Fourier transform of its indicator function at arbitrary spatial frequencies, without any aliasing. Also, it relies on the fact that, on a finite field of view, the Fresnel kernel acts as a low pass filter, which implies in turn that only a limited number of spatial frequencies are necessary to achieve accurate computations of the diffraction pattern in the telescope aperture plane. 

These two aspects combined made it possible to obtain accurate diffraction patterns in about 1hr of GPU time, compared to several days for a direct, 2D FFT based method\cite{aime2024}. We note that the algorithm presented here, based on known methods to compute the continuous Fourier transform of polygon indicator function, retains the advantages of Boundary Diffraction Wave algorithms (in the sense that the most computationally intensive 2D integrals are transformed into manageable 1D contour integrals), as well as those of the traditional Fresnel kernel convolution method, that separate the occulter dependent parts from the (wavelength dependent) Fresnel kernels; this last point might be of relevance if further optimisation on the occulter's shape is conducted beyond the azimuthally averaged transmission profile.

\bibliography{polyFT} 
\bibliographystyle{spiebib} 

\end{document}